\documentclass[12pt]{iopart}

\usepackage{graphicx}
\usepackage{bm}
\usepackage{iopams}
\usepackage{dsfont}
\usepackage[export]{adjustbox}

\usepackage{cite}
\usepackage[colorlinks=true,citecolor=myblue,linkcolor=myblack]{hyperref}

\newtheorem{observation}{Observation}

\newcommand{\slow}{{\rm slow}}
\newcommand{\irr}{{\rm irr}}

\newcommand{\D}{{\rm{d}}}
\newcommand{\x}{\bm{x}}

\newcommand\uleip{Institut f\"ur Theoretische Physik, Universit\"at Leipzig,  Postfach 100 920, D-04009 Leipzig, Germany}

\newcommand\chau{Charles University, Faculty of Mathematics and Physics,
Department of Macromolecular Physics, V Hole{\v s}ovi{\v c}k{\' a}ch 2, 
 CZ-180~00~Praha, Czech Republic}
 
\newcommand\icfo{ICFO-Institut de Ci{\` e}ncies Fot{\` o}niques, The Barcelona Institute of Science and Technology, 08860 Castelldefels (Barcelona), Spain}

\newcommand\geneva{Department of Applied Physics, University of Geneva, 1211 Geneva, Switzerland}

\newcommand\exeter{Department of Physics and Astronomy, University of Exeter, Stocker Road, Exeter EX4 4QL, United Kingdom}

\newcommand\oxford{Department of Materials, University of Oxford, Parks Road, Oxford OX1 3PH, United Kingdom}

\newcommand\potsdam{Institut f\"ur Physik und Astronomie, University of Potsdam, 14476 Potsdam, Germany}

\usepackage{colortbl}
\definecolor{myblue}{rgb}{0.2,0.2,0.8}
\definecolor{myblack}{rgb}{0,0,0}
\definecolor{darkgreen}{rgb}{0,0.6,0.2}

\usepackage{xcolor}

\begin{document}

\title[]{Thermodynamics and optimal protocols of multidimensional quadratic Brownian systems}

\author{Paolo Abiuso$^{1,2}$, 
Viktor Holubec$^3$, 
Janet Anders$^{4,5}$, 
Zhuolin Ye$^6$, 
Federico Cerisola$^{4,7}$, 
and Martí Perarnau-Llobet$^2$}

\address{$^1$ \icfo}
\address{$^2$ \geneva}
\address{$^3$ \chau}
\address{$^4$ \exeter}
\address{$^5$ \potsdam}
\address{$^6$ \uleip}
\address{$^7$ \oxford}

\ead{paolo.abiuso@icfo.eu}

\begin{abstract}
We characterize finite-time thermodynamic processes of multidimensional quadratic overdamped systems.
Analytic expressions are provided for heat, work, and  
dissipation for any evolution of the system covariance matrix. 
The Bures-Wasserstein metric between covariance matrices naturally emerges as the local quantifier of dissipation.
General principles of how to apply these geometric tools to identify optimal protocols are discussed. 
Focusing on the 
relevant slow-driving limit, we show how these results can be used to analyze cases in which the experimental control over the system is partial.
 
\bigskip

\bigskip

\end{abstract}

\maketitle










\section{Introduction}

The minimisation of dissipation  is a central goal in finite-time thermodynamics~\cite{Andresen1984therm,Andresen1984thermII,Deffner2020thermcontrol}. 
In most applications, one is interested in finding the optimal 
time variation of some control parameters, e.g. magnetic or electric fields, in order to achieve a desired task while minimising the amount of energy dissipated to the environment.  
Such tasks could range from the design of a  cycle  for a thermal engine~\cite{seifert2012stochastic,myers2022quantum} 
to the erasure of information in an information processing device~\cite{Zulkowski2014,Proesmans2020,Zhen2021}. 
Finding  optimal protocols in finite time is however often a very challenging task, as it requires a functional optimisation over all possible paths in the control parameter space, 
as well as a perfect understanding of the non-equilibrium dynamics resulting from such control.
In the regime of small mesoscopic systems,  remarkable progress on this topic has been achieved in the last decades with the development of the field of stochastic thermodynamics~\cite{jarzynski1997nonequilibrium,sekimoto1998langevin,crooks1999entropy,seifert2005entropy,jarzynski2011equalities,seifert2012stochastic,miller2017entropy}. 
Optimal drivings are nowadays known for  overdamped~\cite{schmiedl2007optimal,Schmiedl2008,Holubec2014,Plata2019,Zhang2020} and underdamped systems~\cite{Gomez-Marin2008,Dechant2017underdamped}, as well as driven single-level quantum dots~\cite{Esposito2010finite}. 
However, such explicit solutions only exist for one-dimensional systems and are, in general, computationally  hard to scale up.

Other solutions are known for situations in which the control parameter varies {\it slowly} compared to the system relaxation time, as the optimisation admits a geometric formulation~\cite{Ruppeiner1995,Salamon1983,Nulton1985quasi,Crooks2007meas,Sivak2012therm,Zulkowski2012geom,Marcus2014optimal,abiuso2020geometric}  and the problem considerably simplifies.
Indeed, the space of control parameters can be endowed with a Riemannian metric in such a way that geodesic paths correspond to minimally dissipative thermodynamic processes. While the geodesic equations might be hard to solve, the important realisation is that the number of coupled equations is given by the number of control parameters, and independent of the size of the system of interest (by comparison, a full out-of-equilibrium solution of the dynamics needs a number of equations that scales exponentially with the number of components of the system). This has enabled finding optimal driving protocols in such regime for complex systems such as a two dimensional Ising model~\cite{Rotskoff2015Opt,Gingrich2016nearopt}, nanomagnets~\cite{Rotskoff2017geom}, and  quantum spin chains~\cite{scandi2019thermodynamic}. Optimal protocols for different classes of slowly driven heat engines have also been developed by such a geometric approach~\cite{abiuso2020optimal,Brandner2020,Miller2020b,Bhandari2020,TerrenAlonso2022,abiuso2020geometric,frim2021optimal,frim2021geometric,Eglinton2022optimal}. 
Besides the slow driving regime, the optimization problem can also be simplified in the opposite, \emph{fast-driving}, regime~\cite{cavina2021maximum,erdman2019maximum,PhysRevResearch.2.033083,Blaber2021}.

Beyond the slow or fast driving limits, a general connection has been established between optimal transport and minimally-dissipative thermodynamic protocols in the overdamped limit  \cite{aurell2011optimal,aurell2012}. This connection was recently exploited to show that the minimal dissipation in any process governed by a Langevin equation is directly related to the $L^2$-Wasserstein distance between initial and final states~\cite{dechant2019thermodynamic} (see also \cite{VanVu2021,Nakazato2021,dechant2021minimum}). However in general full control on the system's Hamiltonian is needed to saturate this bound. To address the relevant case of partial experimental control, one therefore requires a different approach, that is able to quantify dissipation on non-optimal trajectories.


In this paper, we study thermodynamic transformations for many-body quadratic overdamped systems. 
We derive general expressions for the flux of work and heat, and we show that the dissipation is governed by the Bures-Wasserstein (BW) distance between covariance matrices, which coincides with the $L^2$-Wasserstein distance between Gaussian distributions of \cite{dechant2019thermodynamic,VanVu2021,Nakazato2021,dechant2021minimum}. 
Our derivation allows for a direct generalisation of the well-known single-body case, studied by Schmiedl and Seifert for a single-particle overdamped system \cite{schmiedl2007optimal}, as well as new insights on the form of optimal drivings. In particular, we provide an integral analytic expression for the dissipation valid for  
\emph{any response trajectory} of the system, not necessarily minimally dissipating. This also naturally enables the study of \emph{partial control}. That is, the situation where the limited number of control parameters does not allow for exploring the whole space of states, so that the fundamental lower bounds of \cite{aurell2011optimal,dechant2019thermodynamic} might not be reachable. This is a common scenario in complex systems, where experimentally only a few degrees of freedom are controllable. In order to illustrate application of our results, and to show the difference between partial and global control, we analyse a system of two interacting particles and a particle confined in a 2-dimensional squeezing potential with different control limitations.



\section{Model: Many-body overdamped spring}
\label{sec:gaussian}

We consider a system of $N$ overdamped Brownian particles described by the position vector $\x$ and mutually interacting via the time-dependent potential 
\begin{eqnarray}
\label{eq:potential}
   V(\x,t) = \frac{1}{2}\x^\intercal K(t) \x\;, 
\end{eqnarray}
or, equivalently, via the force field $F(\x) = - \nabla V(\x) = -K \x$ (when possible, we omit writing the time argument from now on). Each particle $i = 1,\dots, N$ might have a different number of degrees of freedom $d_i$, i.e.  $\x \in \mathbb{R}^M$, where $M=\sum_{i=1}^N d_i$. The potential (\ref{eq:potential}) accounts for both self-energy of the individual particles and interactions between the particles.
The stiffness matrix $K$ is symmetric and positive definite $K \geq 0$ (that is, the potential is confining).
Assuming that all the particles have the same friction coefficient $\gamma$, the system dynamics obeys the set of Langevin equations~\cite{risken1996fokker}
\begin{eqnarray}
\gamma\dot{\x} = -K \x + \sqrt{2\gamma k_B T}{\bm \eta}\;,
\label{eq:dxK}
\end{eqnarray}
where the Gaussian noise ${\bm \eta}$ obeys $\left<{\bm \eta}\right> = 0$, its components $\left<\eta_i(t)\eta_j(t')\right>= \delta_{ij}\delta(t-t')$, and $T$ is the temperature of the thermal environment, which we assume is fixed throughout (isothermal). 
From now on, we will use natural units in which $\gamma=1$, $k_B=1$. For the 
general case in which different particles have different friction coefficients, see Sec.~\ref{sec:generalizations}.

Departing from an arbitrary normalized initial distribution, 
the state of the system at time $t$ is represented by a Gaussian probability density function (PDF)~\cite{risken1996fokker}~\footnote{The PDF describing the state under the dynamics (\ref{eq:dxK}) is always Gaussian given Gaussian initial conditions, or after a negligible initial transient \cite{risken1996fokker}.} 
\begin{eqnarray}
 p(\x,t) = \frac{1}{\sqrt{(2\pi)^N \det \Sigma}}
  {\exp}\left(-\frac{\x^\intercal \Sigma^{-1}(t) \x}{2}\right).
  \label{eq:PDFxN}
\end{eqnarray}
Here $\Sigma(t) = \left<\x \x^\intercal\right>(t)$ denotes the covariance matrix at time $t$. The PDF~(\ref{eq:PDFxN}) has zero mean $\left<\x \right>(t)=0$ 
(see Sec.~\ref{sec:generalizations} and App.~\ref{app:general_case} for the more general case). 

The distribution $p(\x,t)$ is therefore defined by its covariance matrix, whose dynamics $\dot{\Sigma} = \left<\x \dot{\x}^\intercal\right> + \left<\dot{\x} \x^\intercal\right>$ can be obtained from Eq.~(\ref{eq:dxK}) and reads~\cite{risken1996fokker} 
\begin{eqnarray}
\label{eq:Sigma_dynamics}
\dot{\Sigma}(t) = -K(t)\Sigma(t) -  \Sigma(t) K(t) + 2 T\;,
\end{eqnarray}
where implicitly $T=T\mathds{1}$.
In case the response dynamics $\Sigma(t)$ is given, and one wants to drive the potential $K(t)$ accordingly (i.e. $K(t)$ is the control protocol generating the response dynamics $\Sigma(t)$), the solution of Eq.~(\ref{eq:Sigma_dynamics}) for $K(t)$ is
\begin{eqnarray}
\label{eq:KgivenSigma}
    K=\int_0^\infty \D \nu\ e^{-\nu\Sigma}(2T-\dot{\Sigma})e^{-\nu\Sigma}=
T \Sigma^{-1}-\int_0^\infty \D\nu\ e^{-\nu\Sigma}\dot{\Sigma}e^{-\nu\Sigma}\;.
\end{eqnarray}
Notice that instantaneous quenches of $K(t)$ can be added at the beginning and at the end of the protocol, without affecting the dynamics of $\Sigma(t)$. For example, to end the transformation in equilibrium, one can add a final quench to $K = T\Sigma^{-1}$.

\paragraph{Remark.} We stress here that the explicit evaluation of Eq.~(\ref{eq:KgivenSigma}) can be performed analytically.
To be consistent with the notation, throughout the text we use the operator $\mathcal{I}(A,B)=\int_0^\infty \D\nu \; e^{-\nu A}B e^{-\nu A}$ expressed in its integral matrix form; at the same time, in the basis that diagonalizes $A$, i.e. $A_{ij}=\delta_{ij}a_i$, the components of this operator can be easily expressed as $\mathcal{I}(A,B)_{ij}=\frac{B_{ij}}{a_i+a_j}$.

\section{Thermodynamics of quadratic systems}
\label{sec:mainmain}

The average energy of a system described by a multidimensional probability distribution~(\ref{eq:PDFxN}) in the potential~(\ref{eq:potential}) reads
\begin{eqnarray}
\label{eq:energy}
    E(t)=\int d\x\; p(\x,t) V(\x,t)
    ={\sum_{ij}\frac{1}{2}K_{ij}(t) \left<x_i x_j\right>(t)}
    =\frac{\Tr[K(t)\Sigma(t)]}{2} \;,
\end{eqnarray}
where 
$\Tr[AB]=\sum_{ij}A_{ij}B_{ji}$. The variation of energy is split canonically in a work contribution, originating in the variation of the external potential, and a heat contribution, stemming from the evolution of the system induced by the dissipative environment~\cite{seifert2012stochastic,sekimoto2010stochastic}.
I.e. the work ($W$) and heat ($Q$) fluxes entering the system are defined as
\begin{eqnarray}
\label{eq:W-Q}
\dot{W}= \frac{\Tr[\dot{K}\Sigma]}{2}\;,\quad \dot{Q}= \frac{\Tr[K\dot{\Sigma}]}{2}\;.
\end{eqnarray}

In a similar fashion to the seminal work by Schmiedl and Seifert~\cite{schmiedl2007optimal} we can 
write the work input of a finite time  transformation of duration $\tau$ as
\begin{eqnarray}\nonumber
    W &= \frac{1}{2}\int_0^\tau dt\;\Tr[\dot{K}\Sigma]\\
    &= \frac{1}{2}\Tr[K\Sigma]\Big|^\tau_0-\frac{T}{2}\log \det{\Sigma}\Big|^\tau_0+\frac{1}{2}\int_0^\tau dt\;\Tr\left[\int_0^\infty d\nu\; e^{-\nu\Sigma}\dot{\Sigma}e^{-\nu\Sigma}\dot{\Sigma}\right]
    \label{eq:multidim_seifert} \;,
\end{eqnarray}
where in the second equality we integrated by parts and used Eq.~(\ref{eq:KgivenSigma}). In the following we will indicate as $\mathcal{Q}(\tau)-\mathcal{Q}(0):=\Delta\mathcal{Q}$ the variation of any quantity $\mathcal{Q}$ during a transformation. Given that $\frac{1}{2}\log\det\Sigma|^\tau_0= -\int d\x\ p \ln p\ |^\tau_0=\Delta S$ is the  variation of Von Neumann entropy of the system~(\ref{eq:PDFxN}), it is possible to rewrite Eq.~(\ref{eq:multidim_seifert}) as
\begin{eqnarray}
\label{eq:work_with_diss}
    W-\left(\Delta E - T \Delta S\right) = \frac{1}{2}\int_0^\tau dt\;\Tr\left[\int_0^\infty d\nu\; e^{-\nu\Sigma}\dot{\Sigma}e^{-\nu\Sigma}\dot{\Sigma}\right]\equiv W_\irr\;,
\end{eqnarray}
This expression identifies the dissipated work, $W_\irr$, of an arbitrary response trajectory~$\Sigma(t)$.
This is our first main result. The above derivation represents a natural multidimensional generalisation of the one-dimensional result of~\cite{schmiedl2007optimal}.

The irreversible work~(\ref{eq:work_with_diss}) turns out to be the integral in time of a quadratic form that coincides with the Bures-Wasserstein (BW) metric on positive-definite matrices~\cite{bengtsson2017geometry,bhatia2019bures}. That is 
$W_\irr=\int_0^\tau dt\; g_\Sigma(\dot{\Sigma},\dot{\Sigma})$\;, where \begin{eqnarray}
\label{eq:BWmetric}
     g_\Sigma(A,B)=\frac{1}{2}\int_0^\infty d\nu\; \Tr[e^{-\nu\Sigma}A e^{-\nu\Sigma}B]\;, \\
     g_\Sigma(d\Sigma,d\Sigma)\equiv D_{\rm BW}(\Sigma,\Sigma+d\Sigma)^2 \;,
\end{eqnarray}
with the latter being the infinitesimal BW squared distance. This metric has been intensely studied  as it appears in problems of statistical inference and metrology in quantum information~\cite{wootters1981statistical,braunstein1994statistical,paris2009quantum,bengtsson2017geometry}, as well as in the theory of optimal transport~\cite{bhatia2019bures,olkin1982distance}. 

For fixed endpoints, the lower bound for $W_\irr$ is obtained for the response trajectory $\bar{\Sigma}(t)$ that minimizes the integral of the quadratic form in Eq.~(\ref{eq:work_with_diss}). That is
 \begin{eqnarray}
 \label{eq:Wmintau}
    W_\irr\geq \frac{D_{\rm BW}(\Sigma_1,\Sigma_2)^2}{\tau}\;,
 \end{eqnarray}
where the BW-geodesic length between the initial and final points $\Sigma(0)=\Sigma_1$, $\Sigma(\tau)=\Sigma_2$, is given by (see \ref{app:geodesics}, or~\cite{bengtsson2017geometry,bhatia2019bures})
\begin{eqnarray}
\label{eq:bures_dist}
    D_{\rm BW}(\Sigma_1,\Sigma_2)^2\equiv \Tr[\Sigma_1]+\Tr[\Sigma_2]-\Tr[\sqrt{\sqrt{\Sigma_1}\Sigma_2\sqrt{\Sigma_1}}]\;.
\end{eqnarray}
The corresponding geodesic, i.e. the minimally dissipating response trajectory, is given by (cf.~\ref{app:geodesics}) 
\begin{eqnarray}
\label{eq:geodesics_BW}
    \bar{\Sigma}(s\tau)=(1-s)^2\Sigma_1+s^2\Sigma_2+s(1-s)\left(\sqrt{\Sigma_1\Sigma_2}+\sqrt{\Sigma_2\Sigma_1}\right)\;,
\end{eqnarray}
with $s=t/\tau$ and thus $0\leq s\leq 1$ independently on the total duration $t$ of the protocol.

The appearance of the distance (\ref{eq:bures_dist}) in (\ref{eq:Wmintau}) is no coincidence: it was realised recently~\cite{dechant2019thermodynamic} that the optimal transport problem is connected to the irreversible entropy production in diffusive dynamics, and its value is minimized by the $L^2$-Wasserstein distance between the initial and final distributions \cite{aurell2011optimal,dechant2019thermodynamic,nakazato2021geometrical}. In the case of Gaussian distributions, the $L^2$-Wasserstein distance coincides with the above BW distance between the covariance matrices (\ref{eq:bures_dist}). 

Besides it being straightforward, one
key advantage of our derivation
is that expression~(\ref{eq:work_with_diss}) is valid for \emph{any} response trajectory and allows to compute $W_\irr$ also when the transformation does not saturate the lower bound~(\ref{eq:Wmintau}). In particular, it can be used for the realistic case of partial experimental control, 
when the system is constrained to explore only a subset of the distributions space (see the following Sec.~\ref{sec:control} and examples in Sec.~\ref{sec:applications}).

\paragraph{Total control vs partial control.}
\label{sec:control}\ \\
In experiments, the system is typically controlled by varying $K(t)$. The optimal control parameter protocol $\bar{K}$  corresponding to the geodesic~(\ref{eq:geodesics_BW}) is determined by substituting $\bar{\Sigma}$ into~(\ref{eq:KgivenSigma}). Perfect implementation of $\bar{K}$ would then saturate the minimal dissipation bound~(\ref{eq:bures_dist}). However, this assumes that $\bar{K}$ is experimentally feasible.
This might not be the case in general. Performing the minimization over a restricted region of control parameters limits the system response to a submanifold of allowed states. In general, this results in a case-dependent minimum value strictly larger than the global minimum, $W_\irr > D_{\rm BW}^2/\tau$, e.g., see Example~\ref{sec:example_RJ} below.

In other cases, the initial and final point of the transformation might not even be fixed (e.g., when optimizing the performance of a cyclic heat engine).  
To show the consequences of fixed/unfixed boundary states $\Sigma(0)$ and $\Sigma(\tau)$, consider that the variation $\dot{\Sigma}=\dot{\Sigma}_d+\dot{\Sigma}_{r}$ can be divided into a diagonal contribution and a non-diagonal, \emph{rotating} contribution. That is, 
given $\Sigma=\sum_i \omega_i |i\rangle \langle i|$, the diagonal part is
$\dot{\Sigma}_d=\sum_i \dot{\omega}_i |i\rangle \langle i|$ and 
the rotating part is $\dot{\Sigma}_r=\sum_i \omega_i ( \dot{|i\rangle} \langle i|+|i\rangle \dot{\langle i|})$.
From Eqs.~(\ref{eq:work_with_diss}) and (\ref{eq:BWmetric}) we know that $ \dot{W}_\irr=g_\Sigma(\dot{\Sigma},\dot{\Sigma})$. It is easy to check that $g_\Sigma(\dot{\Sigma}_d,\dot{\Sigma}_r)=0$ 
which implies 
that the irreversible work naturally decouples into a diagonal and a rotating contribution:
\begin{eqnarray}
    \dot{W}_\irr=g_\Sigma(\dot{\Sigma}_d,\dot{\Sigma}_d)+g_\Sigma(\dot{\Sigma}_r,\dot{\Sigma}_r)\equiv \dot{W}^{(d)}_\irr+\dot{W}^{(r)}_\irr\;.
\end{eqnarray}
Both $\dot{W}^{(d)}_\irr$ and $\dot{W}^{(r)}_\irr$ are positive, which means that the dissipation generated in a non-commuting transformation for $\Sigma$ ($\dot{W}^{(r)}_\irr>0$) is always larger than the commuting case ($\dot{W}^{(r)}_\irr=0$). (A similar phenomenon occurs
for quantum systems, described by their density matrices \cite{abiuso2020geometric}).
At the same time, for any transformation $\Sigma(t)=\sum_i \omega_i(t)|i(t)\rangle\langle i(t)|$, the change in system entropy $\propto\Delta \left[\log \det \Sigma\right]$ and energy $\propto \Delta \left[\Tr[K\Sigma]\right]$ can also be achieved 
by a similar transformation $\Sigma^*(t)=\sum_i \omega_i(t)|i(0)\rangle\langle i(0)|$  in which the covariance matrix commutes with itself at all times $[\Sigma^*(t),\Sigma^*(t')]=0$, and $W_\irr^{(r)}=0$. 
This leads to the following observation:
\begin{observation}
\label{obs:no_rotations}
If the restrictions on the control parameters $K(t)$ allow and $\left[\Sigma(0), \Sigma(\tau)\right]=0$, rotation of the covariance matrix should be avoided in order to minimize dissipation.
\end{observation}
In the fully commuting case, it is easy to see that  Eqs.~(\ref{eq:energy}-\ref{eq:work_with_diss}) simplify and we recover ($K$ being diagonal in the same basis of $\Sigma$, with eigenvalues $k_i$)
\begin{eqnarray}
\nonumber
    E=& \frac{1}{2}\sum_i k_i\omega_i\;,\quad \Delta S= \frac{1}{2}\sum_i \Delta\left[ \log\omega_i\right]\;,\\
    W_\irr=& \int_0^\tau dt\; \sum_i \frac{1}{4} \frac{\dot{\omega}^2_i}{\omega_i}\geq \frac{1}{\tau}\sum_i\left(\sqrt{\omega_i(\tau)}-\sqrt{\omega_i(0)}\right)^2\;.
    \label{eq:Wirr_commuting}
\end{eqnarray}
When reduced to a single mode, this is exactly the result found by Schmiedl and Seifert in~\cite{schmiedl2007optimal},  which we thus see being \emph{extensive in the eigenmodes} $\omega_i$ of $\Sigma$: that is, all the modes $\{\omega_i,k_i\}$ can be treated as effectively independent in the commuting case. As an instance of a transformation with fixed boundaries that force non-commutation, see Example \ref{sec:example_rot}.

As mentioned above, controlling the potential~(\ref{eq:potential}) in time defines the evolution of the state via the dynamical equation~(\ref{eq:Sigma_dynamics}). Conversely, a given response trajectory $\Sigma(t)$ is translated to its generating control $K(t)$ through Eq.~(\ref{eq:KgivenSigma}).  This means that fixing the boundary values of $\Sigma$ is non-trivially related to fixing boundary controls.  The results of optimisation thus strongly depend on the imposed constraints~\cite{ye2022optimal}. 
At the same time, for the purpose of typical applications to isothermal processes (cf. Sec.~\ref{sec:applications}), in which the goal is to minimize work dissipation, we can consider the \emph{slow-driving} limit of the dynamics~\cite{slowdriving}. In this limit the potential $K(t)$ is modified slowly, more precisely we assume %
$\dot{K} \sim 1/\tau$ with
$\tau$ much larger than the relaxation timescale of the system $\tau\gg \gamma/|K|$, and it is sufficient to solve the dynamical equation~(\ref{eq:Sigma_dynamics}) up to the first order in $1/\tau$. The zeroth order corresponds to the quasistatic limit $\tau\rightarrow \infty$. This allows to expand
any state-dependent quantity $\mathcal{Q}$ 
around its equilibrium value $\mathcal{Q}^{(0)}$, keeping only the leading correction term $\mathcal{Q}^{(1)}\sim\mathcal{O}(1/\tau)$. In our specific setting, the
covariance matrix can be expanded as
\begin{eqnarray}
    \Sigma(t)=\Sigma^{(0)}(t)+\Sigma^{(1)}(t)+\mathcal{O}(1/\tau^2)
\end{eqnarray}
with $\Sigma^{(0)}=T K^{-1}$ and $    \Sigma^{(1)} =-T\int_0^\infty ds\; e^{-s/K}\frac{d}{dt}{(K^{-1})}e^{-s/K}$ (cf. Eqs.~(\ref{eq:Sigma_dynamics}) and~(\ref{eq:KgivenSigma})).
However,
the irreversible work~(\ref{eq:work_with_diss}) is already of order $\mathcal{O}(1/\tau)$. 
To express the dissipation in the slow regime, it
is therefore sufficient to substitute $\Sigma^{(0)}$ in~(\ref{eq:work_with_diss}). 
In other words, we observe that

\begin{observation}
\label{obs:SD_equiv}
In the  slow-driving limit, controlling the inverse stiffness matrix $TK^{-1}(t)$ of the potential is equivalent to directly steering the covariance matrix $\Sigma(t)$. 
The irreversible work in the slow-driving limit therefore reads
\begin{eqnarray}
W_\irr^\slow=\frac{T}{2}\int_0^\tau dt\;
g_{K^{-1}}\left(\frac{d}{dt}(K^{-1}),\frac{d}{dt}(K^{-1})\right)
\;.
\end{eqnarray}
\end{observation}

\section{Generalizations}
\label{sec:generalizations}
In the previous section, we have focused on the paradigmatic case of the potential~(\ref{eq:potential}) and 
density distributions~(\ref{eq:PDFxN}) 
centered around $\x=0$, and a particle-independent friction coefficient.
Nevertheless, the obtained results
are fully extendable also when removing such assumptions. 

First, in~\ref{app:general_case} we solve the general case of a quadratic potential with time-dependent center $\bm{z}(t)$, i.e. $V(\x,t)=\frac{1}{2}({\bm x-\bm z(t)})K(t)({\bm x-\bm z(t)})$. As the system is in general driven out of equilibrium, the center of the potential does not necessarily coincide with the average particle position, $\langle\x\rangle\equiv \bm \xi(t)\neq \bm z(t)$, and the irreversible work gains an additional contribution  (see details in~\ref{app:general_case}). 
Focusing on the limit of slow driving, it can be expressed as
\begin{eqnarray}
\label{eq:Wirr_moving}
W_\irr^\slow& = \int_0^\tau dt\; \left(|\dot{\bm \xi}|^2 + g_{\Sigma}(\dot{\Sigma},\dot{\Sigma})\right)\;.
\end{eqnarray}
Similarly to (\ref{eq:bures_dist}), the lower bound for $W_\irr^\slow$ is in this case
\begin{eqnarray}
\label{eq:lowerbound_withmu}
   W_\irr^\slow\geq \frac{1}{\tau} \left(|\bm \xi_1 -\bm \xi_2|^2 + D_{BW}(\Sigma_1,\Sigma_2)^2 \right)\;.
\end{eqnarray}
Observations~\ref{obs:no_rotations} and \ref{obs:SD_equiv} from Section~\ref{sec:mainmain}
remain valid: if possible, rotations
of the covariance matrix thus should be avoided;
the state variables in the expression~(\ref{eq:Wirr_moving}) can be substituted by their equilibrium values $(\bm \xi, \Sigma)=(\bm z,TK^{-1})$.
Moreover, in the same limit, $\bm\xi(t)-\bm z(t)\sim \mathcal{O}(\tau^{-1})$ while the associated correction to quasistatic $\Delta E$ and $\Delta S$ is negligible $\mathcal{O}(\tau^{-2})$ (cf. \ref{app:general_case}). This implies that moving the trap $\dot{\bm\xi}\neq 0$ only contributes to the dissipation (\ref{eq:lowerbound_withmu}) and should therefore be avoided when possible, in the same spirit as Observation~\ref{obs:no_rotations}.


Second, we comment on the generalization to systems where different particles have different friction coefficients $\gamma_i$.
In such a case, the Langevin 
equations (\ref{eq:dxK}) become, in components,
\begin{eqnarray}
    \gamma_i \dot{x}_i = -\sum_j K_{ij}x_j + \sqrt{2\gamma_i k_B T}\eta_i \;.
\end{eqnarray}
Notice that some of the $\gamma_i$ might refer to different degrees of freedom of the same particle. 
The white noises $\eta_i$ are mutually uncorrelated.
We define 
$
y_i\equiv\sqrt{\gamma_i} x_i
$
and rewrite the Langevin equations as
\begin{eqnarray}
\label{eq:dxKy}
\dot{\bm y}=-K'\bm y + \sqrt{2 k_B T} \bm \eta,
\end{eqnarray}
where the transformed stiffness matrix $K'_{ij}=\frac{K_{ij}}{\sqrt{\gamma_i\gamma_j}}$ 
is still symmetric and positive definite. At the same time, the covariance matrix of the $\bm y$ variable, $\Sigma'= \left<\bm y \bm y^\intercal\right>$, satisfies
$\Sigma'_{ij}=\sqrt{\gamma_i}\Sigma_{ij}\sqrt{\gamma_i}\;.$
Finally 
the energy of the system is given by
\begin{eqnarray}
\label{eq:energy-y}
    E=\frac{1}{2}\Tr[\Sigma K] =\frac{1}{2}\Tr[\Sigma' K']\;,
\end{eqnarray}
and similarly $\dot{Q}=\frac{1}{2}\Tr[\dot{\Sigma}'K']$ and $\dot{W}=\frac{1}{2}\Tr[\Sigma'\dot{K}']$. Given the formal equivalence between Eqs.~(\ref{eq:dxK},\ref{eq:energy},\ref{eq:W-Q}) and (\ref{eq:dxKy},\ref{eq:energy-y}) above, 
we see that the problem is equivalently mapped to the case with fixed $\gamma_i=\gamma$, $\forall i$.

Finally, throughout the paper we assumed a fixed temperature $T$. At the same time, the expressions for energy~(\ref{eq:energy}), heat and work~(\ref{eq:W-Q}), as well as $\Delta S=\frac{1}{2}\Delta\log\det\Sigma$ do not intrinsically depend on the temperature $T$. We can therefore relax such assumption and admit a time-dependent temperature $T(t)$ \cite{brandner2015thermodynamics,brandner2016periodic}. In such case the definition of irreversible work becomes 
\begin{eqnarray}
W_\irr= 
W-\left(\Delta E- \int_0^\tau dt\; T \dot{S}\right)=-Q+ \int_0^\tau dt\; T \dot{S}=
\int_0^\tau dt\; T \dot{S}_\irr
\;,
\end{eqnarray}
$ \dot{S}_\irr$ being the irreversible entropy production. From the derivation in Section~\ref{sec:mainmain} we get the same expression $W_\irr=\frac{1}{2}\int_0^\tau dt\;g_{\Sigma}(\dot{\Sigma},\dot{\Sigma})$, as well as the validity of all the above observations and generalizations.

\section{Applications}
\label{sec:applications}
Here, we present  two examples of application of the formalism, results and observations  introduced above.

\subsection{Interacting particles in double trap}
\label{sec:example_RJ}

\begin{figure}
     \qquad\qquad\quad\ \ (a)    \includegraphics[width=0.4\textwidth,valign=t]{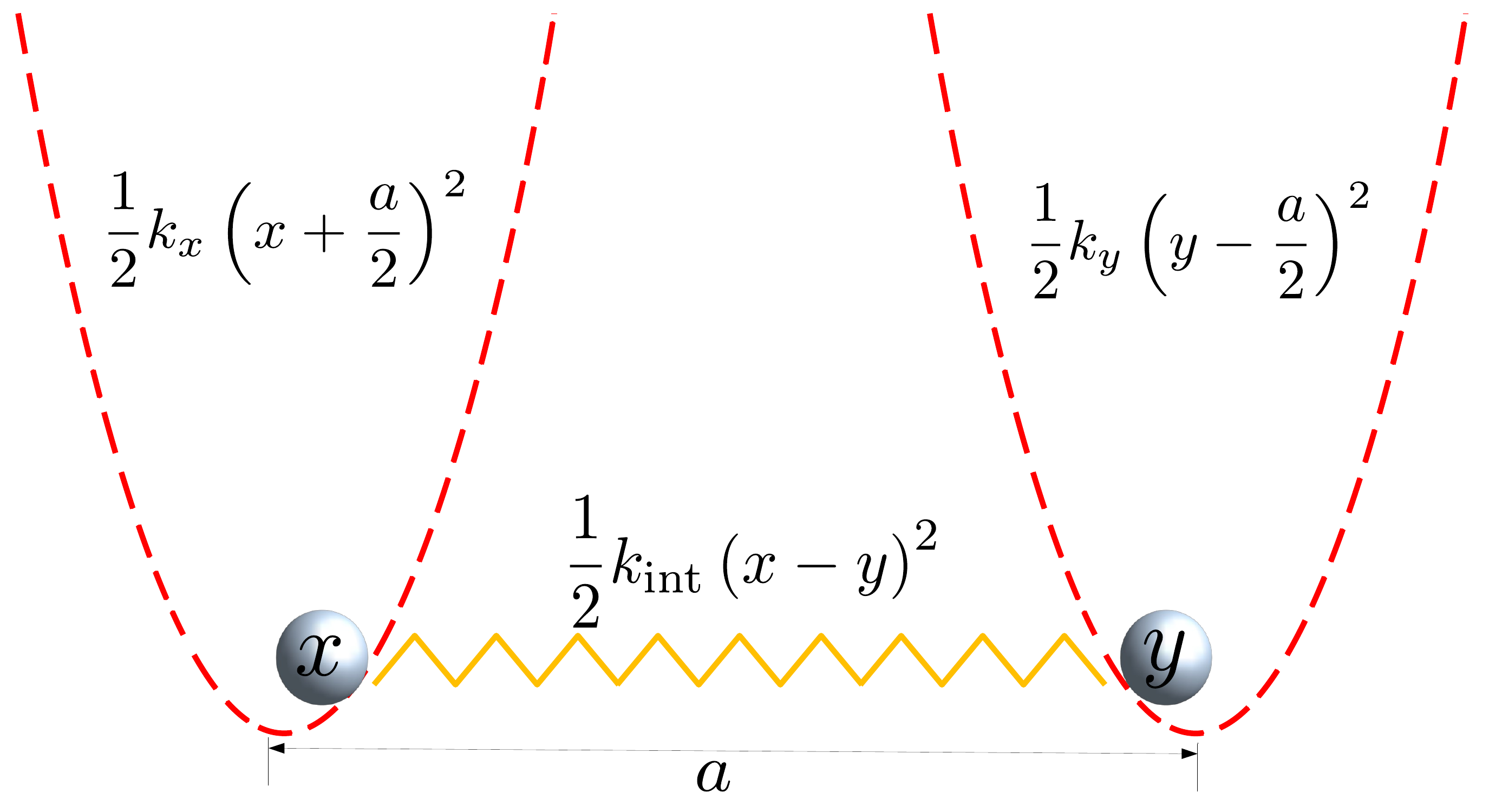}
     \\
\flushright
     (b)     \includegraphics[width=0.8\textwidth,valign=t]{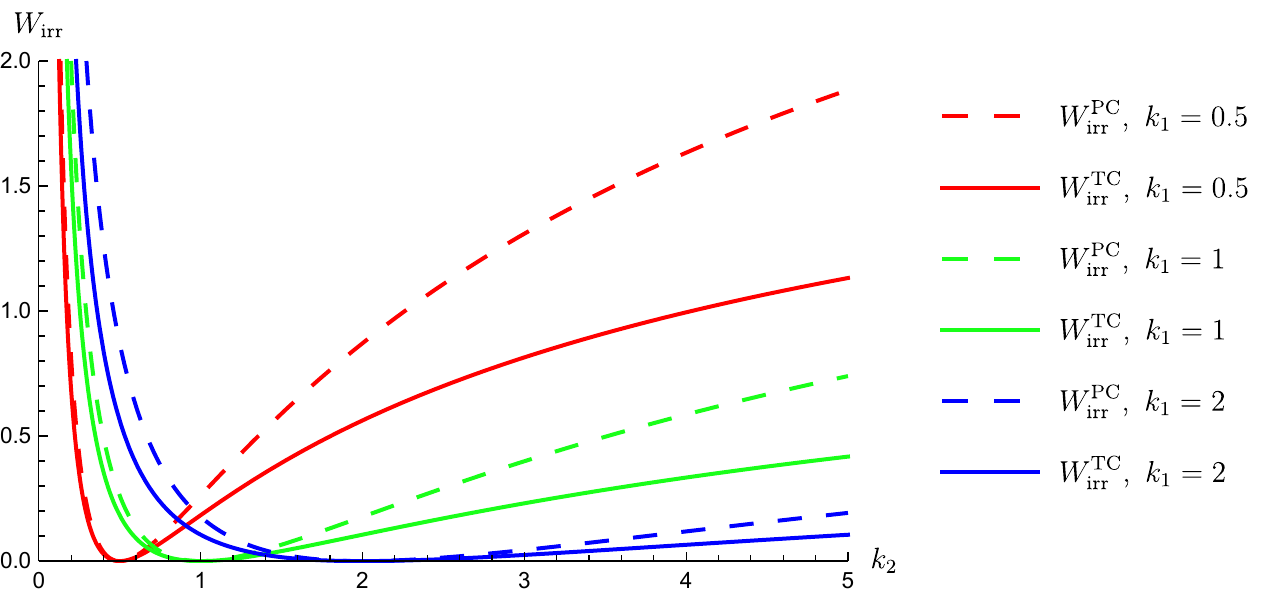}
\caption{(a) Two particles, each locally confined in a harmonic potential, with stiffnesses $k_x=k_y$, and an additional harmonic interaction, $k_{\rm int}$, between them. A transformation is performed to modify the local traps stiffnesses.
(b) Plots show the minimum values of the dissipated work $W_\irr$ for transformations from $k_x(0)=k_y(0)=k_1$ to $k_x(\tau)=k_y(\tau)=k_2$ as a function of $k_2$, and for various values of $k_1$. 
$W^{\rm PC}_\irr$ is the minimum work dissipation attainable with \emph{partial control}, i.e. when modifying only $k_x$ and $k_y$ between the end points. $W^{\rm PC}_\irr$ is always larger than its corresponding \emph{total control} counterpart, $W^{\rm TC}_\irr$. The latter is attainable when controlling also $k_{\rm int}$ and $a$ {\it during} the transformation, while returning them to their initial values, $k_{\rm int} (\tau) = k_{\rm int}(0)$ and $a(\tau) = a(0)$.
Both $k_1$ and $k_2$ are given in units of $k_{\rm int}(0)=1$. The remaining parameters, $T$, $a(0)$ and $\tau$, are set to 1.}
    \label{fig:double_trap}
\end{figure}

First, we show how partial control over a system can substantially increase the amount of dissipation when compared to the optimal geodesics transformation. Consider the case of two particles on a line, \emph{Romeo and Juliet}, who are constrained to be located at two different places, separated by a distance $a$. That is, Romeo (Juliet) is at position $x$ ($y$) and subject to a confining harmonic potential of strength $k_x$ ($k_y$) centered at $\frac{a}{2}$ ($-\frac{a}{2}$). At the same time, the two particles feel a harmonic attraction of strength $k_{\rm int}$. The complete system is described by the potential  (cf. Fig.~\ref{fig:double_trap})
\begin{eqnarray}
    V=\frac{1}{2}k_x\left(x-\frac{a}{2}\right)^2+\frac{1}{2}k_y\left(y+\frac{a}{2}\right)^2+\frac{1}{2}k_{\rm int}\left(x-y\right)^2\;,
\label{eq:RJ_potential}
\end{eqnarray}
For two colloidal particles, such an interaction can be realized using optical tweezers~\cite{jones2015optical} or an effective potential induced by feedback control~\cite{Khadka2018}. Besides, it qualitatively mimics the interaction of trapped active particles studied in Ref.~\cite{krishnamurthy2021synergistic} or in a similar model~\cite{mamede2021obtaining}.

Now imagine that an experimenter can operate a transformation of the Hamiltonian parameters with the goal to increase the strength of the local traps, but with a minimal energetic cost. That is, the boundaries of the transformation are  $k_x(0)=k_y(0)<k_x(\tau)=k_y(\tau)$ while  $a(0)=a(\tau)$ and $k_{\rm int}(0)= k_{\rm int}(\tau)$. We want to know the minimum dissipation that an  experimenter can achieve for such a transformation.
In the~\ref{app:example1}, we derive 
and compare the minimum dissipation protocols in the slow-driving limit for two paradigmatic cases: \emph{i) partial control} in which the experimenter can only tune the values of $k_x$ and $k_y$ (while $a$ and $k_{\rm int}$ are both constant), \emph{ii) total control} in which the experimenter can control $k_{\rm int}$ and $a$ as well. The comparison among the two situations is presented in Fig.~\ref{fig:double_trap}. As expected, the
dissipation under partial control, $W_\irr^{\rm PC}$, is always larger than that for total control, $W_\irr^{\rm TC}$. In particular we observe that having the possibility of controlling all the parameters of the potential~(\ref{eq:RJ_potential}) allows, in general, substantial savings of up to $\simeq 50\%$ of energy dissipation with respect to simply tuning the stiffnesses $k_{x,y}$.

\begin{figure}
    \centering
    (a)
    \includegraphics[width=0.6\textwidth,valign=t]{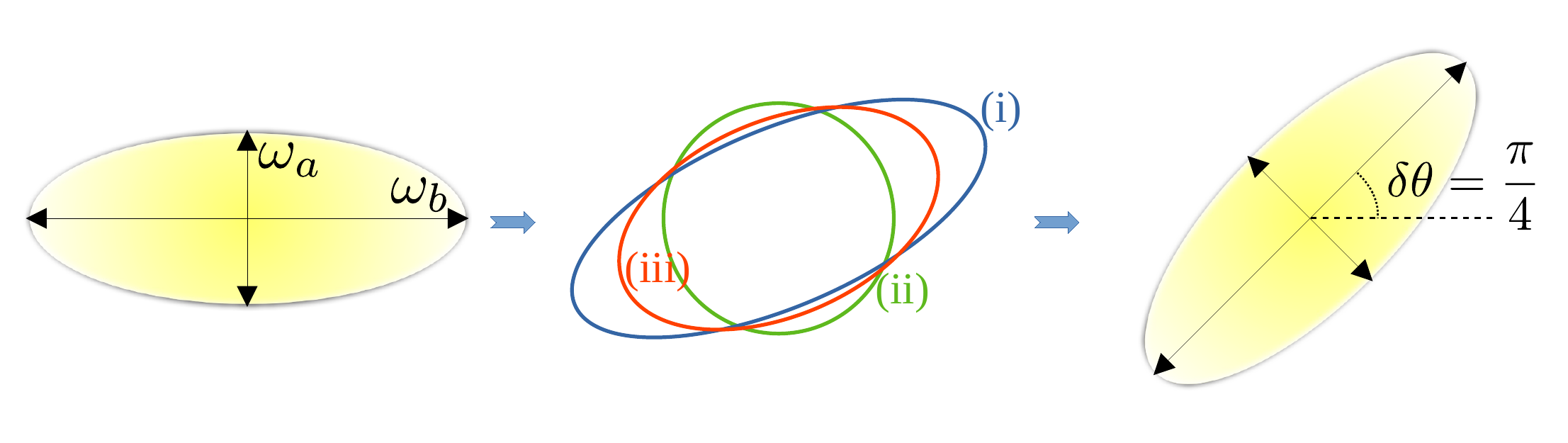}\\
    (b)
    \includegraphics[width=0.6\textwidth,valign=t]{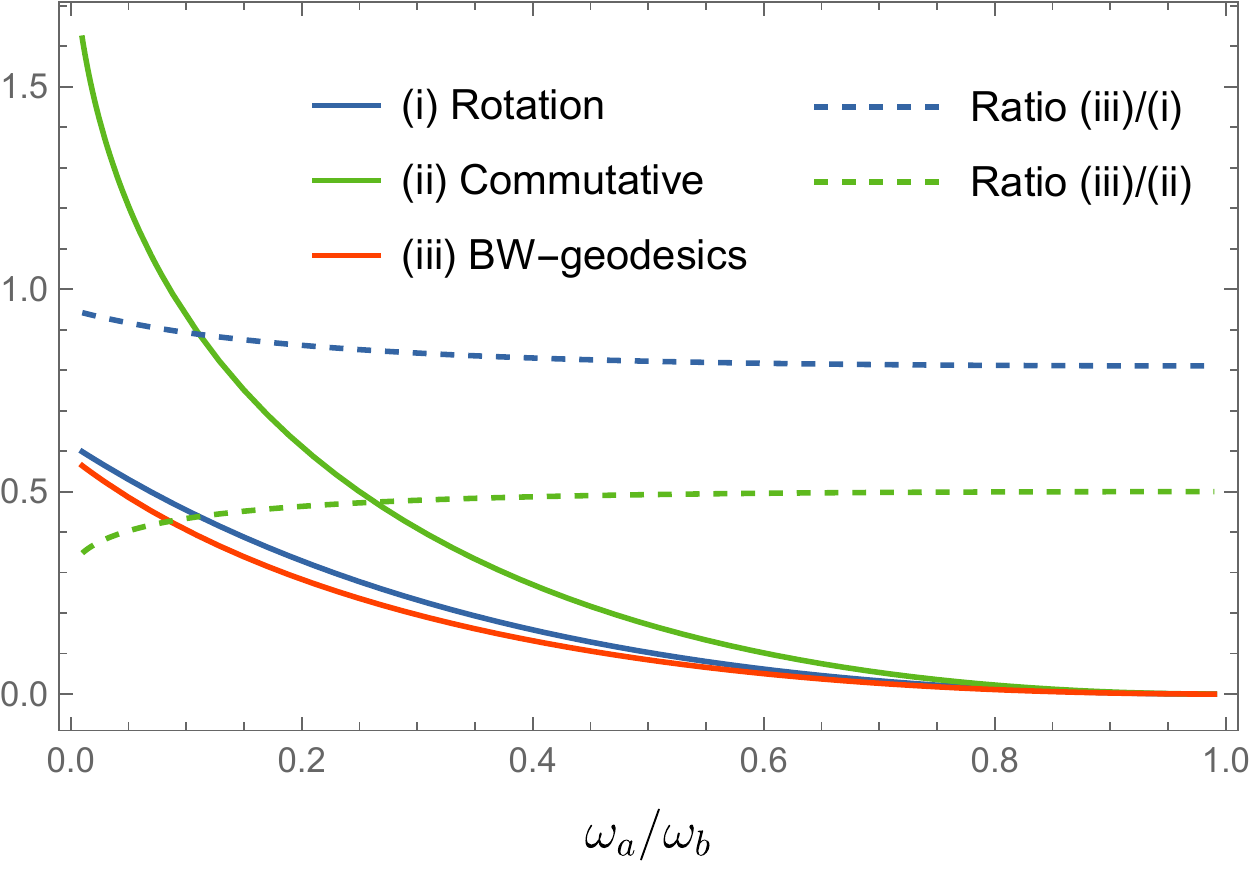}
    \caption{
    (a) 
    Strategies introduced in the text for the rotation of a 2-dimensional Gaussian system by the
    angle $\delta\theta=\frac{\pi}{4}$:
    (i) uniform rotation of the system
    (blue), (ii) commutatively squeezing the system until it is isotropic and, subsequently, stretching it in the desired direction (green), (iii) following the optimal BW-geodesics (red). 
    (b) The continuous lines show the 
    $W_\irr$ 
    in units of $\omega_b/\tau$ 
    for the different strategies. The dashed lines depict
    the normalised comparison between the different strategies. The commutative strategy (i) is highly dissipative, while the rotation strategy (ii) is comparable to the optimal protocol, dissipating just $\sim 20\%$ more energy than (iii) for most values of $\omega_a/\omega_b$.
    }
    \label{fig:2D_rotation}
\end{figure}

\subsection{Rotating a 2-dimensional squeezed potential}
\label{sec:example_rot}
As a second example, we consider the rotation of a two-dimensional Gaussian system in the $xy$ plane. 
Specifically, we consider a Gaussian PDF with a non-isotropic covariance matrix of
the position coordinates $\{x,y\}$
\begin{eqnarray}
    \Sigma=\pmatrix{
    \left< x^2 \right> &  \left< xy \right> \cr
     \left< xy \right> &  \left< y^2 \right>
    }\;,
\end{eqnarray}
which is 
squeezed with the major axis 
and the $x$-axis forming an angle $\theta$ 
in the $xy$ plane.  
Denoting the eigenvalues of $\Sigma$ as $\omega_a$ and $\omega_b$, it can be written in the form
\begin{eqnarray}
    \Sigma_{\theta,\omega_a,\omega_b}=
    \pmatrix{
    \cos^2(\theta)\omega_a+\sin^2(\theta)\omega_b & \sin(\theta)\cos(\theta)(\omega_a-\omega_b) \cr
    \sin(\theta)\cos(\theta)(\omega_a-\omega_b) & \sin^2(\theta)\omega_a+\cos^2(\theta)\omega_b 
    } \;.
\end{eqnarray}
Our goal is to find a minimum-dissipation protocol that would overall rotate the system in the $xy$ plane, by an angle $\delta\theta=\frac{\pi}{4}$.
We thus consider protocols starting at $\Sigma(0)=\Sigma_{0,\omega_a,\omega_b}$ and ending at $\Sigma(\tau)=\Sigma_{\frac{\pi}{4},\omega_a,\omega_b}$ (cf. Fig.~\ref{fig:2D_rotation}), and compare three
strategies for accomplishing this task: 
(i) Uniform rotation trajectory: the experimenter simply rotates the system, i.e., fixes 
the variances $\omega_{a,b}$ in $\Sigma$ and increases $\theta$;
(ii) Commutative trajectory: the experimenter tunes $\omega_{a,b}$ to the same value at an intermediate time, thus making the distribution isotropic. Afterwards, they re-stretch the distribution in the desired direction. Such protocol is commutative in the sense that $[\dot{\Sigma}(t),\Sigma(t)]=0\ \forall t$ (notice however $[{\Sigma}(0),\Sigma(\tau)]\neq 0$). The intermediate point can be chosen optimally to minimize the dissipation;
(iii) Optimal protocol: the experimenter is able to control the system such that it follows the BW-geodesic $\bar{\Sigma}(t)$  in~(\ref{eq:geodesics_BW}) between $\Sigma_{0,\omega_a,\omega_b}$ and $\Sigma_{\frac{\pi}{4}\omega_a,\omega_b}$.
Details of the calculations for each of the trajectories are given in \ref{app:example2} and the results are depicted in Fig.~\ref{fig:2D_rotation}.
We find that the commutative strategy is strongly non-optimal, and dissipates at least twice as much as the geodesics trajectory. Notice that this is not in contradiction to Observation~\ref{obs:no_rotations}, as non-commuting boundary condition induce, in general, non-commuting optimal trajectories.
At the same time we see that the uniform rotation of the system $\Sigma$ is close
to the optimal (geodesic) trajectory in terms of dissipation, while being much simpler to implement.
Finally, no timescale approximation was used in this case, but we notice that in the slow-driving limit the above strategies are equivalently translated to the stiffness matrix $K=T\Sigma^{-1}$ (cf. Observation~\ref{obs:SD_equiv}).

\section{Discussion}
In this paper, we have studied the work, heat exchange, and irreversible work dissipation of overdamped multidimensional classical systems. 
These may have an arbitrary number of degrees of freedom and are confined by harmonic potentials whose parameters can be partially or totally controlled. 
Such systems are described by multidimensional Gaussian probability distributions~\cite{risken1996fokker}.
For uniform friction and non-moving trap centers, 
we have derived a general analytic  expression~(\ref{eq:work_with_diss}) for the irreversible work ($\propto$ entropy production). This expression is  valid for any response trajectory, and allows geometric optimisation based on the Bures-Wasserstein metric for positive matrices.
We also discussed straightforward generalizations of these results to non-uniform friction values and nontrivial trap center dynamics.
Given that in the slow-driving limit there is a one-to-one mapping between the set of reachable states $\Sigma$ and the set of reachable controls $K$, this further allows optimization of control protocols that incorporate experimental constraints, i.e. partial control. 
Finally, we described general design principles for optimal parameter protocols that minimise dissipation 
and illustrated them for two examples, increasing local confinement of two interacting particles and rotating a squeezed potential.


The obtained results point towards the manageable optimization of control protocols in experimental systems with many degrees of freedom~\cite{hucul2008transport,renaut2013chip,Khadka2018,Mayer2020,krishnamurthy2021synergistic}, and they can be directly applied to  optical tweezers setups and electric circuits  \cite{jones2015optical,ciliberto2017experiments,gao2017optical}, 
that wish to minimise dissipation by choosing optimised control parameter protocols.  
Moreover, our findings can be readily applied to the case of engines and refrigerators described in the low-dissipation regime, characterized by the $1/\tau$ scaling of dissipation \cite{schmiedl2007optimal,esposito-EMP-bounds,Martinez2016,abiuso2020geometric,ma2018universal,Ye2021}. 
Further extensions include the analysis of underdamped classical systems, as well as that of quantum Gaussian systems.


\ack
VH gratefully acknowledges support by the Humboldt foundation and by the Czech Science Foundation (project No. 20-02955J).
PA is supported by “la Caixa” Foundation (ID 100010434, Grant No. LCF/BQ/DI19/11730023), and by the Government of Spain (FIS2020-TRANQI and Severo Ochoa CEX2019-000910-S), Fundacio Cellex, Fundacio Mir-Puig, Generalitat de Catalunya (CERCA, AGAUR SGR 1381. 
FC gratefully acknowledges funding from the Fundational Questions Institute Fund (FQXi-IAF19-01).
JA gratefully acknowledges funding from EPSRC (EP/R045577/1).
ZY is grateful for the sponsorship of China Scholarship Council (CSC) under Grant No. 201906310136. 
MPL acknowledges financial support from the Swiss National Science Foundation through an Ambizione grant PZ00P2-186067.

\medskip

\section*{References}

\bibliography{BIB.bib}

\appendix

\section{On the Bures-Wasserstein distance}
\label{app:geodesics}
The Bures-Wasserstein (BW) distance can be defined between positive semidefinite matrices $\Sigma\geq 0$, and its infinitesimal value is given by the metrics $g_{\Sigma}$
\begin{eqnarray}
    D_{\rm BW}(\Sigma,\Sigma+d\Sigma)^2= g_\Sigma(d\Sigma,d\Sigma)=\frac{1}{2}\int_0^\infty ds\; \Tr[e^{-s\Sigma}d\Sigma e^{-s\Sigma}d\Sigma]\;.
\end{eqnarray}
When applied to complex positive matrices of unit trace (that is, \emph{states} in the field of quantum information), such metric represents a fundamental quantifier in problems of quantum statistical inference and metrology~\cite{wootters1981statistical,braunstein1994statistical,paris2009quantum,bengtsson2017geometry}. At the same time it has its own relevance as a distance quantifier between positive real matrices or multivariate distributions, in the context of optimal transport theory~\cite{bhatia2019bures,olkin1982distance}.
The integrated geodesics length between two points $\Sigma_1$ and $\Sigma_2$, if no constraints are imposed on the trace of the matrices, reads 
\begin{eqnarray}
    D_{\rm BW}(\Sigma_1,\Sigma_2)=\sqrt{\Tr[\Sigma_1]+\Tr[\Sigma_2]-\Tr[\sqrt{\sqrt{\Sigma_1}\Sigma_2\sqrt{\Sigma_1}}]}
\end{eqnarray}
and the corresponding geodesics is
\begin{eqnarray}
    \Sigma(t)=(1-t)^2\Sigma_1+t^2\Sigma_2+t(1-t)\left(\sqrt{\Sigma_1\Sigma_2}+\sqrt{\Sigma_2\Sigma_1}\right)
\end{eqnarray}
where the square root
\begin{eqnarray}
    \sqrt{\Sigma_1\Sigma_2}=
    \Sigma_1^{\frac{1}{2}}(\Sigma_1^{\frac{1}{2}}\Sigma_2 \Sigma_1^{\frac{1}{2}})^{\frac{1}{2}}\Sigma_1^{-\frac{1}{2}}
\end{eqnarray}
is the only matrix $R$ satisfying $R^2=\Sigma_1\Sigma_2$ and having a positive spectrum (cf. \cite{bhatia2019bures}).

\section{General case and slow-driving solution}
\label{app:general_case}
Here we consider the case in which also the first moment of the quadratic potential can be driven. 
We assume~\cite{risken1996fokker} that the state is Gaussian with covariance matrix $\Sigma(t)$ and first moments ${\bm \xi(t)}$ (we avoid expliciting time when possible)
\begin{equation}
 p(\x,t) = \frac{1}{\sqrt{(2\pi)^N \det \Sigma}}
  {\exp}\left(-\frac{(\x-{\bm \xi})^\intercal \Sigma^{-1} (\x-{\bm \xi}))}{2}\right)\;,
\end{equation}
while the potential is
\begin{equation}
    V(\x,t)=\frac{1}{2}({\bm x-\bm z (t)})K(t)(\bm x-\bm z (t))\;,
\end{equation}
with $\bm \xi \neq \bm z$ in general. The energy of the system is therefore
\begin{eqnarray} \nonumber
    E(t) &=\int d\x\; p(\x,t)V(\x,t)\\ &=\frac{1}{2}\Tr[\Sigma K]+\frac{1}{2}({\bm \xi-\bm z})K({\bm \xi-\bm z})=\frac{1}{2}\Tr[\Sigma_z K]\;,
\end{eqnarray}
where $\Sigma_z$ is the covariance matrix \emph{centered in $\bm z$}, that is
\begin{eqnarray}
    \Sigma_z=\langle (\x -\bm z)(\x -\bm z)^\intercal\rangle =
    \Sigma+(\bm \xi -\bm z)(\bm \xi -\bm z)^\intercal\;.
\end{eqnarray}
The Langevin equation~(\ref{eq:dxK}) becomes accordingly $\dot{\x}=-K(\x -\bm{z})+\sqrt{2T}\bm{\eta}$ in natural units, which is translated on the Gaussian moments as
\begin{eqnarray}
\label{eq:moment_dynamics_gen}
    \dot{\bm \xi} &=-K(\bm\xi-\bm z)\;, \\
    \partial_t{\Sigma}_z &=-K\Sigma_z-\Sigma_z K+2T\;,
    \label{eq:sigmaZ_dyn}
\end{eqnarray}
where the partial derivative in time is due to the fact that $\Sigma_z$ depends as well on $\bm z$, i.e.
\begin{eqnarray}
\dot{\Sigma}_z=\partial_t\Sigma_z- \dot{\bm z}(\bm \xi-\bm z)^\intercal-(\bm \xi- \bm z)\dot{\bm z}^\intercal\;.
\end{eqnarray}
The work and heat 
can be computed by simply taking the derivative w.r.t. the driving parameters $K$, $\bm z$ (for the work), and the dynamical parameters $\Sigma$, $\bm \xi$ (for the heat), i.e.
\begin{eqnarray}
\nonumber
    W &=\frac{1}{2}\int_0^\tau dt\; \left( \Tr[\dot{K}\Sigma_z]+\partial_{\bm z}\Tr[K\Sigma_z] \dot{\bm z}\right)\\ 
\nonumber
    &=\frac{1}{2}\Tr[K\Sigma_z]\Big|^\tau_0-\frac{1}{2}\int_0^\tau dt\; \left(\Tr[K\dot{\Sigma}_z]-\partial_{\bm z}\Tr[K\Sigma_z] \dot{\bm z}\right)\\
    &=\frac{1}{2}\Tr[K\Sigma_z]\Big|^\tau_0-\frac{1}{2}\int_0^\tau dt\; \Tr[K\partial_t{\Sigma}_z]\;.
\end{eqnarray}
Using the same steps as 
in the main text (integration by parts and Eq.~(\ref{eq:sigmaZ_dyn})), this expression translates to
\begin{eqnarray}\nonumber
    W=& \frac{1}{2}\Tr[K\Sigma_z]\Big|^\tau_0 +
    \frac{T}{2}\int_0^\tau dt\; \Tr[\Sigma_z^{-1}\partial_t\Sigma_z]+\\
    &-\frac{1}{2}\int_0^\tau dt\; \Tr[\int_0^\infty ds\ e^{-s\Sigma_z}\partial_t \Sigma_z e^{-s\Sigma_z}\partial_t\Sigma_z]\;,
\end{eqnarray}
which can be rewritten as
\begin{eqnarray}\nonumber
    W-\Delta E - \frac{T}{2}\Delta \det{\Sigma}_z=\\
    =\frac{T}{2}\int_0^\tau dt \Tr[\Sigma_z^{-1}(\dot{\Sigma}_z-\partial_t\Sigma_z)]
   + \int_0^\tau dt\; g_{\Sigma_z}(\partial_t{\Sigma_z},\partial_t \Sigma_z)\;,
   \label{eq:Wirrnot}
\end{eqnarray}
with the BW metrics (\ref{eq:BWmetric}) $g$.
Notice that in general $\frac{1}{2}\Delta \det \Sigma_z\neq \frac{1}{2}\Delta\det \Sigma =\Delta S$ and therefore the expression above cannot be identified as 
the irreversible work. At the same time, 
minimizing dissipation requires using finite time protocols in which the system ends 
in equilibrium with the thermal bath, so that no dissipation follows 
the end of the protocol. This is automatically satisfied in the case of slow-protocols (see below). For general transformations, it is sufficient to add a final quench of the controls, 
$(K(\tau),\bm z(\tau))=(T\Sigma_z^{-1}(\tau),\bm \xi (\tau))$. The condition $\bm \xi (\tau)=\bm z (\tau)$ is sufficient to rewrite (\ref{eq:Wirrnot}) as
\begin{eqnarray}
   W_\irr=\frac{T}{2}\int_0^\tau dt \Tr[\Sigma_z^{-1}(\dot{\Sigma}_z-\partial_t\Sigma_z)]
   + \int_0^\tau dt\; g_{\Sigma_z}(\partial_t{\Sigma_z},\partial_t \Sigma_z)\;,
\end{eqnarray}
which can be computed explicitly using $
    \partial_t\Sigma_z-\dot{\Sigma}_z = \dot{\bm z}(\bm \xi-\bm z)^\intercal + (\bm \xi- \bm z)\dot{\bm z}^\intercal $.

\subsection{The slow case}
In the slow-driving regime a first order expansion is performed around $\frac{1}{\tau}\simeq 0$ \cite{slowdriving}. For example in the quasistatic limit of $\tau\rightarrow\infty$ the solution for the dynamics~(\ref{eq:moment_dynamics_gen},\ref{eq:sigmaZ_dyn}) is clearly $\bm\xi^{(0)}=\bm z$ and $\Sigma_z^{(0)}=\Sigma^{(0)}=TK^{-1}$. 
The finite time expansion leads to
\begin{eqnarray}\nonumber
    \bm \xi= \bm \xi^{(0)} + \bm \xi^{(1)} + \bm \xi^{(2)}+\dots\quad + \bm\xi^{(i)}\sim\mathcal{O}(\tau^{-i})\\
    \bm \xi^{(0)}=\bm z\;,\quad \bm \xi^{(1)}=-K^{-1}\dot{\bm z}\;.
\end{eqnarray}
As 
$\bm \xi-\bm z=-K^{-1}\dot{\bm z}+ \mathcal{O}(\tau^{-2})$,  
we also get
\begin{eqnarray}
\Sigma_z=\Sigma+ \dot{\bm z}K^{-2}\dot{\bm z}^\intercal  +\mathcal{O}(\tau^{-3})\;,\\
 \dot{\Sigma}_z-\partial_t\Sigma_z= \{\dot{\bm z}\dot{\bm z}^\intercal, K^{-1}\}+ \mathcal{O}(\tau^{-3})=  \{\dot{\bm \xi}\dot{\bm \xi}^\intercal, K^{-1}\}+ \mathcal{O}(\tau^{-3})\;.
\end{eqnarray}
Using the above expressions, the irreversible work 
reads
\begin{eqnarray}
W_\irr& = \int_0^\tau dt\; \left(|\dot{\bm \xi}|^2 + g_{\Sigma}(\dot{\Sigma},\dot{\Sigma})\right)\;+\mathcal{O}(\tau^{-2}) \\
&= \int_0^\tau dt\; \left(|\dot{\bm z}|^2 + T g_{K^{-1}}(\dot{K^{-1}},\dot{K^{-1}})\right)\;+\mathcal{O}(\tau^{-2})\;.
\end{eqnarray}



\section{Detailed and solved examples} 
\label{app:applications}

\subsection{Double trap}
\label{app:example1}

Consider the potential
\begin{eqnarray}
    V=\frac{1}{2}k_x\left(x-\frac{a}{2}\right)^2+\frac{1}{2}k_y\left(y+\frac{a}{2}\right)^2+\frac{1}{2}k_{int}\left(x-y\right)^2\;,
\end{eqnarray}
which can be rewritten in matrix form as
\begin{eqnarray}
    V=\frac{1}{2}(\x -\bm a)^\intercal K (\x-\bm a)+\frac{1}{2}\x^\intercal K_{\rm int}\x
    \end{eqnarray}\,
with
\begin{eqnarray}
    \x=\pmatrix{ x \cr y
    },
    \bm a=\pmatrix{ a/2\cr -a/2
    },
    K=\pmatrix{ k_x & 0\cr 0 & k_y
    },
    K_{\rm int}=\pmatrix{ k_{\rm int} & -k_{\rm int}\cr -k_{\rm int} & k_{\rm int}
    }.
\end{eqnarray}
To use Eq.(\ref{eq:Wirr_moving}), we rewrite the potential in the ``canonical form''
\begin{eqnarray}\nonumber
    V=& \frac{1}{2}(\x-\bm a')^\intercal (K+K_{\rm int}) (\x-\bm a')+\\ &-\frac{1}{2}\bm a'^\intercal (K+K_{\rm int})\bm a'+\frac{1}{2}\bm a^\intercal K \bm a ,
\end{eqnarray}
where 
\begin{eqnarray}
\label{eq:a'}
    \bm a'=(K+K_{\rm int})^{-1}K \bm a
\end{eqnarray}
is the effective center of the potential. The scalar $-\frac{1}{2}\bm a'^\intercal (K+K_{\rm int})\bm a'+\frac{1}{2}\bm a^\intercal K \bm a $ is just a global shift in energy that does not depend on the dynamics of the system and vanishes for cyclic protocols. 

\paragraph{Confining the particles - Irreversibility parameter.} We compute the irreversible work using the slow-driving approximation Eq.(\ref{eq:Wirr_moving}), in which the center of the distribution can be substituted by the center of the potential, and the covariance matrix can be substituted by the inverse stiffness matrix (cf. \ref{app:general_case}), leading to
\begin{eqnarray}
\label{eq:Wirr_gapp}
W_\irr = \int_0^\tau dt\; \left(|\dot{\bm a'}|^2 +  g_{\Sigma}(\dot{\Sigma},\dot{\Sigma})\right)\quad {\rm with} \quad\Sigma=T(K+K_{\rm int})^{-1}\;.
\end{eqnarray}
Suppose the experimenter wants increase the strength of the local traps to increase the confinement of the two particles. The endpoint of the transformation will therefore be
\begin{eqnarray}
a(0)=a(\tau)\;,\  k_{\rm int}(0)= k_{\rm int}(\tau)\;,\ k_x(0)=k_y(0)<k_x(\tau)=k_y(\tau)\;.
\end{eqnarray}
We notice that due to the symmetry of the potential at the boundary, the eigenvectors of $K+K_{\rm int}$ are always $(1,1)$ and $(1,-1)$, independently of the values of $k_{\rm int}$ and $k_x=k_y$. That is, $[(K+K_{\rm int})(0),(K+K_{\rm int})(\tau)]=0$ and we can therefore assume that it commutes with itself at all times (cf. Observation\ref{obs:no_rotations}). In such case, the contribution of $g_{\Sigma}(\dot{\Sigma},\dot{\Sigma})$ to $W_\irr$ simplifies to (cf. Eq.(\ref{eq:Wirr_commuting}))
\begin{eqnarray}
    g_{\Sigma}(\dot{\Sigma},\dot{\Sigma})=\frac{1}{2}\int_0^\tau dt\;\Tr\left[\int_0^\infty ds\; e^{-s\Sigma}\dot{\Sigma}e^{-s\Sigma}\dot{\Sigma}\right] =\frac{1}{4}\left(\frac{\dot{\omega}_1^2}{\omega_1}+\frac{\dot{\omega}_2^2}{\omega_2}\right)\;,
\end{eqnarray}
where $\omega_i$ are the eigenvalues of $\Sigma=T(K+K_{\rm int})^{-1}$, which are easily computed. In particular given $k_x=k_y\equiv k$ we have
\begin{eqnarray}
   \omega_1=\frac{T}{k}\;,\quad
   \omega_2=\frac{T}{k+2k_{\rm int}}\;,
\end{eqnarray}
and therefore
\begin{eqnarray}
    g_{\Sigma}(\dot{\Sigma},\dot{\Sigma})=\frac{T}{4}\left(\frac{\dot{k}^2}{k^3}+\frac{(\dot{k}+2\dot{k}_{\rm int})^2}{(k+2k_{\rm int})^3} \right)\;.
    \label{eq:gSSdotSdot}
\end{eqnarray}
The contribution $|\dot{\bm a}'|^2$ to $W_\irr$ follows from Eq.~(\ref{eq:a'}):
\begin{eqnarray}
   \bm a'=& \frac{k}{k+2k_{\rm int}} \frac{a}{2} \pmatrix{1 \cr -1} \\
   |\dot{\bm a}'|^2 = & 2 a^2 \left(\frac{k}{k+2k_{\rm int}} \right)^4 \left(\frac{d}{dt}\left(\frac{k_{\rm int}}{k}\right)\right)^2 + \frac{\dot{a}^2}{2}\left(\frac{k}{k+2k_{\rm int}}\right)^2\;.
   \label{eq:a'dot2}
\end{eqnarray}
The associated dissipation of the transformation can be computed from the expressions above for any slow transformation.
We now consider the \emph{partial control} (PC) case in which the distance $a$ and interaction strength $k_{\rm int}$ is fixed, and the experimenter can only control the local stiffnesses 
$k_x=k_y\equiv k$. Substituting (\ref{eq:gSSdotSdot}) and (\ref{eq:a'dot2}) with $\dot{a}=\dot{k}_{\rm int}=0$, the irreversible work (\ref{eq:Wirr_gapp}) then specifies to
\begin{eqnarray}
    \dot{W}_\irr =\frac{T}{4}\left( \frac{1}{k^3}+ \frac{1}{(k+2k_{\rm int})^3} \right)\dot{k}^2 + 2a^2 \frac{k^2_{\rm int}}{(k+2k_{\rm int})^4}\dot{k}^2\;.
    \label{eq:disPC}
\end{eqnarray}
For fixed boundary values of $k$, it can be proven using the Cauchy-Schwarz inequality that the dissipation with partial control~(\ref{eq:disPC}) is lower-bounded by
\begin{eqnarray}
   W^{\rm PC}_\irr\geq \frac{1}{\tau}\left(\int_{k_1}^{k_2}\D k\; 
   \sqrt{
   \frac{T}{4}\left( \frac{1}{k^3}+ \frac{1}{(k+2k_{\rm int})^3} \right) + 2a^2 \frac{k^2_{\rm int}}{(k+2k_{\rm int})^4}
   }
   \right)^2\;.
\end{eqnarray}
By comparison, in the case of \emph{total control} (TC), the bound for 
the dissipation is given by (\ref{eq:lowerbound_withmu}), which, in our case, reads 
\begin{eqnarray}
    W^{\rm TC}_\irr\geq \frac{1}{\tau}\left(|\bm a'_1 - \bm a'_2|^2 +D_{\rm BW}(\Sigma_1,\Sigma_2)^2\right)\;,\quad {\rm with}\\
   |\bm a'_1 - \bm a'_2|^2=\frac{a^2}{2}\left(\frac{k_1}{k_1+2k_{\rm int}}-\frac{k_2}{k_2+2k_{\rm int}}\right)^2\;,\\
   \frac{D_{\rm BW}(\Sigma_1,\Sigma_2)^2}{T}=\left(\frac{1}{\sqrt{k_{1}}}-\frac{1}{\sqrt{k_{2}}}\right)^2 +\left(\frac{1}{\sqrt{k_{1}+2k_{\rm int}}}-\frac{1}{\sqrt{k_{2}+2k_{\rm int}}}\right)^2  \;.
\end{eqnarray}

\subsection{Rotating a 2-dimensional system}
\label{app:example2}
In this section, we consider the rotation of a covariance matrix $\Sigma$ in 2 dimensions by an angle $\theta=\frac{\pi}{4}$. 
We thus impose the boundary conditions
\begin{eqnarray}
\label{eq:sigma_in_fin}
\Sigma_{\rm in}=\pmatrix{
    \omega_a & 0 \cr 0 & \omega_b
} \;, 
\qquad 
\Sigma_{\rm fin}=\frac{1}{2}\pmatrix{
    \omega_a+\omega_b & \omega_a-\omega_b \cr \omega_a-\omega_b & \omega_a+\omega_b
}   \;.
\end{eqnarray}
And we minimize the irreversible work~(\ref{eq:work_with_diss}) according to three possible strategies.

\paragraph{(i). Simple rotation protocol}

First, we consider the transformation from $\Sigma_{\rm in}$ to $\Sigma_{\rm fin}$ to be performed by 
uniformly rotating the experimental apparatus, without modifying the squeezing $\{\omega_a,\omega_b\}$ of the distribution. This corresponds to an angle-parametrized protocol
\begin{eqnarray}
    \Sigma_\theta=\pmatrix{
    \cos^2(\theta)\omega_a+\sin^2(\theta)\omega_b    & \sin(\theta)\cos(\theta)(\omega_a-\omega_b) \cr
    \sin(\theta)\cos(\theta)(\omega_a-\omega_b) & \sin^2(\theta)\omega_a+\cos^2(\theta)\omega_b
}
\end{eqnarray}
starting at $\Sigma_0\equiv\Sigma_{\rm in}$ and ending at 
$\Sigma_{\frac{\pi}{4}}\equiv\Sigma_{\rm fin}$.
The irreversible work production~(\ref{eq:work_with_diss}) is in this case
given by
$
W_\irr=\frac{1}{2}\int_0^\tau dt\;\Tr\left[\int_0^\infty d\nu\; e^{-\nu\Sigma}\dot{\Sigma}e^{-\nu\Sigma}\dot{\Sigma}\right]
$.
Given the rotational symmetry of the problem, it is obvious that the optimal rotation of the system will have a constant 
speed $\dot{\theta}$. Thus the 
integrand
\begin{eqnarray}
\label{eq:integrandapp}
\dot{W}_\irr=\frac{1}{2}\Tr\left[\int_0^\infty d\nu\; e^{-\nu\Sigma}\dot{\Sigma}e^{-\nu\Sigma}\dot{\Sigma}\right]
\end{eqnarray}
will be constant in time and can be computed, e.g. for $\theta=0$, which yields 
\begin{eqnarray}
\Sigma=\pmatrix{
    \omega_a & 0 \cr 0 & \omega_b
}   \;, \quad
\dot{\Sigma}=\dot{\theta}
\pmatrix{0 &  \omega_a-\omega_b \cr
        \omega_a-\omega_b & 0} \;.
\end{eqnarray}
Now we use the fact that the operator $\mathcal{I}(A,B)=\int_0^\infty \D\nu \; e^{-\nu A}B e^{-\nu A}$ can be easily expressed in components as $\mathcal{I}(A,B)_{ij}=\frac{B_{ij}}{a_i+a_j}$, in the basis that diagonalizes $A$, i.e. $A_{ij}=\delta_{ij}a_i$. We therefore get
\begin{eqnarray}
\int_0^\infty \D\nu \; e^{-\nu \Sigma}\dot{\Sigma} e^{-\nu \Sigma}=\dot{\theta} \frac{\omega_a-\omega_b}{\omega_a+\omega_b}\pmatrix{0 & 1 \cr 1 & 0}\;,
\end{eqnarray}
from which it is easy to compute the value of (\ref{eq:integrandapp}):
\begin{eqnarray}
\dot{W}_\irr=\dot{\theta}^2\frac{(\omega_a-\omega_b)^2}{\omega_a+\omega_b}\;.
\end{eqnarray}
The
minimum value of ${W}_\irr=\int_0^\tau\dot{W}_\irr$ for the uniform rotation over the 
total angle $\Delta \theta=\frac{\pi}{4}$ is thus given by
\begin{eqnarray}
W_\irr=\frac{1}{\tau}\frac{\pi^2}{16}\frac{(\omega_a-\omega_b)^2}{\omega_a+\omega_b}\;.
\end{eqnarray}

\paragraph{(ii). Commutative protocol}
One possible way to interpolate between $\Sigma_{\rm in}$ and $\Sigma_{\rm fin}$ (\ref{eq:sigma_in_fin}) is to change the values of $\omega_{a,b}$ to reach an intermediate symmetric covariance matrix
\begin{eqnarray}
\Sigma_{\rm intermediate}=\pmatrix{\omega_c & 0 \cr 0 & \omega_c}\;,
\end{eqnarray}
which is proportional to the identity matrix, and later ``re-stretch it" in the $\pi/4$ direction in the same way. Such protocol is at all times commutative, in the sense that $[\Sigma(t),\dot{\Sigma}(t)]=0\ \forall t$, although the final and initial covariance matrices do not commute. The total irreversible work for such a strategy is clearly twice the irreversible work obtained for the transformation $\Sigma_{\rm in}\rightarrow\Sigma_{\rm intermediate}$, in a time $\tau/2$. We therefore get, using the commutative result (\ref{eq:Wirr_commuting}),
\begin{eqnarray}
W_\irr=\frac{4}{\tau}\left((\sqrt{\omega_a}-\sqrt{\omega_c})^2+(\sqrt{\omega_b}-\sqrt{\omega_c})^2\right)\;.
\end{eqnarray}
$\omega_c$ is a free parameter of the described protocol, which can be chosen to minimize $W_\irr$. 
The optimal choice is $\omega_c=\left(\frac{\sqrt{\omega_a}+\sqrt{\omega_b}}{2}\right)^2$, leading to the 
minimum 
dissipation for commutative protocols
\begin{eqnarray}
W_\irr=\frac{2}{\tau}(\sqrt{\omega_a}-\sqrt{\omega_b})^2\;.
\end{eqnarray}

\paragraph{(iii). Optimal protocol}
The minimal value of dissipation for any protocol between $\Sigma_{\rm in}$ and $\Sigma_{\rm fin}$ is given by the main lower bound~(\ref{eq:Wmintau}), which is saturated when performing the BW-geodesics~(\ref{eq:geodesics_BW}).
In our case, we obtain
\begin{eqnarray}
\nonumber
    \frac{D_{\rm BW}(\Sigma_{\rm in},\Sigma_{\rm fin})^2}{\tau}&=\frac{1}{\tau} \left(\Tr[\Sigma_{\rm in}]+\Tr[\Sigma_{\rm fin}]-\Tr[\sqrt{\sqrt{\Sigma_{\rm in}}\Sigma_{\rm fin}\sqrt{\Sigma_{\rm in}}}]\right)\\
    &=\frac{1}{\tau}\left( 2(\omega_a+\omega_b)-\sqrt{2}\sqrt{2(\omega_a+\omega_b)^2-(\omega_a-\omega_b)^2}\right)\;.
\end{eqnarray}

\end{document}